# Exploring the Use of Drones for Taking Accessible Selfies with Elderly

**Yuan Yao**
Tsinghua University
Beijing, China
yaoyuan18@mails.tsinghua.edu.cn

**Callum Parker**
The University of Sydney
Sydney, Australia
callum.parker@sydney.edu.au

**Weiwei Zhang**
Tsinghua University
Beijing, China
zww19@mails.tsinghua.edu.cn

**Jihong Jeung**
Tsinghua University
Beijing, China
jihong95@tsinghua.edu.cn

**Soojeong Yoo**
The University of Sydney
Sydney, Australia
soojeong.yoo@sydney.edu.au

## Abstract
Selfie taking is a popular social pastime, and is an important part of socialising online. This activity is popular with young people but is also becoming more prevalent with older generations. Despite this, there are a number of accessibility issues when taking selfies. In this research, we investigate preferences from elderly citizens when taking a selfie, to understand the current challenges. As a potential solution to address the challenges identified, we propose the use of drones and present a novel concept for hands free selfie taking. With this work, we hope to trigger conversation around how such a technology can be utilised to enable elderly citizens, and more broadly people with physical disabilities, the ability to easily take part in this social pastime.

## Introduction
Selfies are self-portrait photographs usually taken with smartphones. Taking selfies is popular social activity with people aged between 21 to 30[1]. As the demographic of technology usage widens, people of all ages are beginning to take selfies to share with their friends and family. Selfies, while a fun type of photo, can also be a useful for socialising - triggering conversation [27, 26], understanding someone's wellbeing [23], and community engagement [13, 12, 30, 9, 16, 32].

| Reference | Year | Main content | User type | Control type | Study design | Study Size |
|---|---|---|---|---|---|---|
| Jane et al. [21] | 2017 | Cross-culture design | Chinese / American | Speech / Gesture | Comparative study | 18 |
| Chien et al.[6] | 2015 | Selfies | Travellers / Individuals | Personal device | No user study | - |
| Jessica et al.[5] | 2016 | Affective Computing | Common people | Autonomous | Evaluation study | 20 |
| Ashley et al.[8] | 2017 | Navigation | Jogger / Walker | Gesture | Field study | 110 |
| Sara et al. [10] | 2019 | Dancing | Artistes | Whole-body | Preliminary study | 3 |
| Avila et al. [2] | 2015 | Navigation | Visually Impaired Persons | Speech / Personal device | Evaluation study | 1 |
| Florian et al. [28] | 2015 | Accompany | Jogger | Whole-body | Evaluation study | 13 |
| Pascal et al. [18] | 2018 | Levitating tangibles | Player | Gesture | Evaluation study | 17 |
| Sara et al. [17] | 2011 | Sports training | Athlete | Whole-body | Preliminary study | 1 |
| Pascal et al. [19] | 2017 | VR Tactile Feedback | Player | Gesture | No user study | - |

**Table 1:** Different application directions and target users of Human-Drone Interaction from 2011-2019.

However, not everyone has photography skills to take an aesthetically pleasing selfie [22]. The current methods to take selfies are also not very inclusive as they can require a lot of stretching to include the user in the frame while capturing the background, which may not be possible for those with physical disabilities and the elderly [29]. To overcome this, a "selfie stick" could be used to help reduce stretching, but it can be cumbersome to carry and difficult to hold for elderly people, particularly those who experience hand tremors[2]. Therefore, more understanding is needed around alternative methods for elderly selfie taking and the uses for the selfies.

Drones have potential to overcome these issues as they allow for photos to be taken from virtually any angle and hands-free operation [15, 11]. Table 1 shows that there has been much research on the interactions between humans and drones, specifically referred to as Human-Drone Interaction (HDI) [3]. This work is often combined with cameras [14, 20, 17], screens and projectors to provide navigation for pedestrians [8], joggers [28], and visually impaired users [2]. At the same time, work has also shown how drones can be used to provide companionship [4, 5, 21].

While all this work has demonstrated the usefulness of drones for extending interaction, more understanding is needed regarding the design of drones for taking selfies with elderly to ensure they are accessible to a wider audience. In response to this, we investigated elderly preferences around selfie taking. From our findings, we present an initial concept for a selfie taking drone designed for elders and outline key opportunities and challenges that fu-ture work should investigate. Ultimately, the aim of this work is to trigger discussion around inclusive design of drones for selfie taking.

---
[1]Selfie City - http://selfiecity.net/#findings
[2]Essential tremors - https://www.betterhealth.vic.gov.au/health/conditionsandtreatments/essential-tremor

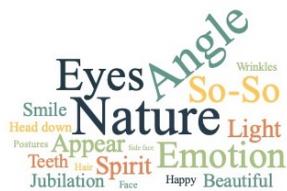

**Figure 1:** Keywords for elderly citizen interview.

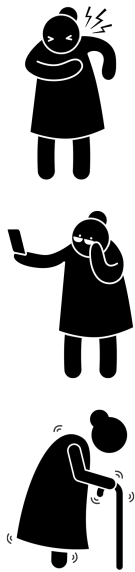

**Figure 2:** Physical difficulties of elderly.

## User Study

*Study Setup*
We organised one-on-one interviews in Nanjing, China with elderly who take selfies and use smartphones. Our workshop in Nanjing was attended by 35 elderly people, ranging in age from 60 to 89 (*mean*=74.9, *SD*=7.95). All of them were retired and the majority had more than one year experience using a smartphone. The main goal of the workshop was to learn about the inconvenience of using a smartphone and preferences around taking selfies. The interviews lasted 10-15 minutes and focused on six questions:

- What problems do you have when using a smartphone?
- How do you evaluate nice photos?
- How much do you like selfies and in which situations do you take them?
- How often do you take selfies?
- What problems do you have with taking selfies?
- How do you evaluate a nice selfie?

*Findings*
We will now summarise the two main findings from our study:

*1.Elderly and Selfie*
During the interview, we found that our elderly participants are passionate about taking photos and selfies. They mentioned that its necessary for them to take more photos and go through old photos in case their memory fades. And most of them prefer the natural photos (Figure 1). Preference for photo subjects include family, pets, and plants. At the same time, they also like taking landscapes photos while travelling. They also expressed a desire to share.

*2.The problem with selfies*
There are two major challenges with taking selfies that were brought up during the interviews. The first is a physical problem, our participants mentioned that they often experienced difficulties with straightening their arms and had trouble keeping a pose for long periods of time to take a selfie(Figure 2). Therefore, it can be difficult to find the right perspective. Additionally, some participants also cited that their hands are shaky making it difficult to keep the camera still - often resulting in blurred photos. The second challenge identified is that most of our elderly participants had some form of Presbyopia(Figure 1) and mentioned that they had trouble seeing their smartphone's screen clearly. Particularly, some apps with small and complex icons are difficult for them. Therefore, they have to rely on others to help them take photos.

In summary, we found that taking photos and selfies is needed by elderly citizens who want to record their lives and share them with their family and friends. However, it is plagued by smartphone use and serious physical problems.

## Concept
To tackle the challenges identified in the previous section, such as stretching, shaking, and visual, we explore how these challenges can be overcome through the design of a selfie taking drone. The purpose is to provide a photography assist function for the elderly who cannot take satisfactory pictures due to physical difficulties. Figure3(a) shows the main functionality of the drone concept, where it can implicitly adjust the distance and composition of the photo based on an individual user's preferences. The selection of distance includes four levels of far away, long distance, medium distance and short distance, which can correspond to long shots, full shots, medium shots, and close-ups in photography language. Furthermore, the perspective of the shot can be adjusted to match the composition.

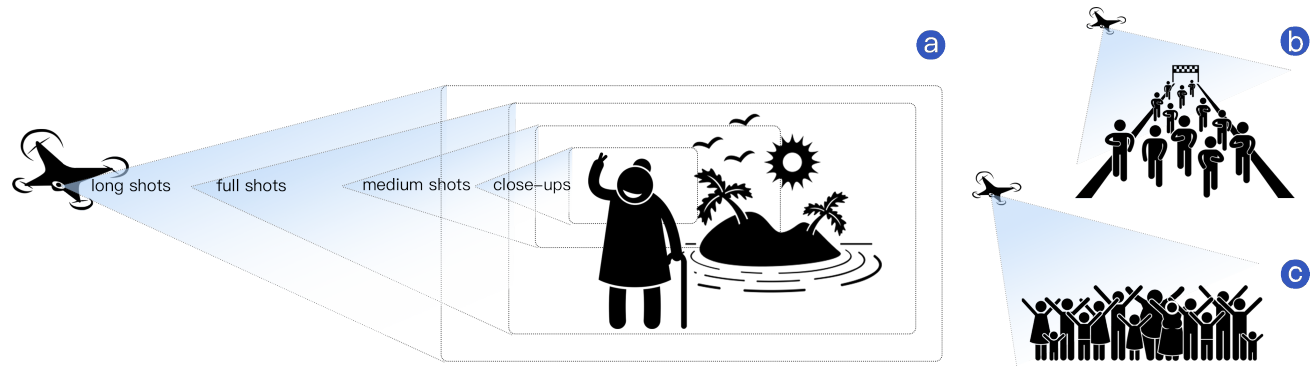

**Figure 3:** (a)Taking photos with different shots and perspectives by drone; (b) Group photo - family; (c) Group photo- event.

Based on the interview data, we identified that elderly enjoy taking group selfies with friends and family (Figure 3(b)(c)). Therefore a drone is useful to ensure everyone fits into the frame by implicitly adjusting its distance and angle through computer vision techniques, such as face or skeletal tracking. Furthermore, in the field of automatic composition research, there are many mature methods to support the interaction of drone selfies, such as it integrates the View Proposal Network (VPN), a deep learning-based model that outputs composition suggestions. [7, 24, 25].

## Discussion and Future work

In this research, we identified key challenges that affect elderly when taking selfies. In response to these challenges we propose a concept of a personalised selfie taking drone. We also found that HDI has an active design space. Especially around photo-shooting. However, as a developing technology, it still has problems that need to be improved. For instance, Table 1 shows that gestures are a popular interface in HDI, but it requires high accuracy and low latency. Therefore the technology needs to be mature. Speech interaction is also common, and enables hands-free control. It can be considered more accessible than other methods [31]. However, in outdoor environments, environmental noise and propeller sound could cause interference. There is also the more conservative method of interacting through a remote controller or smartphone. However, this may present a learning burden to the elderly.

Future work should continue to investigate the design of accessible drone interfaces and further explore the poten-tial of selfie taking drones with the elderly, with a particular focus on exploring a interaction methods and feedback (visual, sound and tactile[1] ).

**Acknowledgement**

This work has been supported by Tsinghua University Scientific Research Foundation project: NO.20197010002, Research on portrait parametric design for elderly users.